\documentclass[12pt]{article}
\usepackage{color}
\usepackage{latexsym}
\usepackage{epsfig,amssymb,euscript, mathrsfs,cite}
\usepackage{amsmath}
\definecolor{MyDarkBlue}{rgb}{0.15,0.15,0.45}
\usepackage[linktocpage=true]{hyperref}
\usepackage{enumerate}
\hypersetup{
colorlinks=true,
citecolor=MyDarkBlue,
linkcolor=MyDarkBlue,
urlcolor=MyDarkBlue,
pdfauthor={},
pdftitle={},
pdfsubject={hep-th}
}
\newsavebox{\ns}
\newsavebox{\dbrane}
\newsavebox{\dbshort}

\def\be{\begin{equation}}
\def\ee{\end{equation}}
\def\bea{\begin{eqnarray}}
\def\eea{\end{eqnarray}}

\newcommand{\nn}{\nonumber}

\newcommand\R{\mathbb{R}}

\newcommand\diff{\mathrm{d}}

\newcommand{\dd}{\mathrm{d}}

\newcommand{\ii}{\mathrm{i}}

\newcommand{\ex}{\mathrm{e}}
\newcommand{\vol}{\mathrm{vol}}

\newcommand{\Imag}{\mathrm{Im}\,}
\newcommand{\Real}{\mathrm{Re}\,}

\newlength{\sswidth}

\newcommand\cN{\mathcal{N}}
\newcommand\tr{\mathrm{Tr}}

\newcommand{\sgn}{\mathrm{sgn}}

\numberwithin{equation}{section}       

\begin{document}

\bibliographystyle{utphys}

\begin{titlepage}

\begin{center}

\today

\vskip 2.3 cm 

\vskip 5mm

{\Large \bf D2-brane Chern-Simons theories: }

\vskip 7mm

{\Large \bf $F$-maximization $=$ $a$-maximization}

\vskip 15mm

{Martin Fluder and James Sparks}

\vspace{1cm}
\centerline{{\it Mathematical Institute, University of Oxford,}}
\centerline{{\it Andrew Wiles Building, Radcliffe Observatory Quarter,}}
\centerline{{\it Woodstock Road, Oxford, OX2 6GG, UK}}

\end{center}

\vskip 2 cm

\begin{abstract}
\noindent We study a system of $N$ D2-branes probing a generic Calabi-Yau three-fold singularity in the presence 
of a non-zero quantized Romans mass $n$. We argue that the low-energy effective $\mathcal{N}=2$ Chern-Simons 
quiver gauge theory flows to a superconformal fixed point in the IR, and construct the dual 
AdS$_4$ solution in massive IIA supergravity. We compute the free energy $F$ of the gauge theory on $S^3$ 
using localization. In the large $N$ limit we find $F=c\, (nN)^{1/3} a^{2/3}$, where 
$c$ is a universal constant and $a$ is the $a$-function of the ``parent'' four-dimensional 
$\mathcal{N}=1$ theory on $N$ D3-branes probing the same Calabi-Yau singularity. 
It follows that maximizing $F$ over the space of admissible R-symmetries 
is equivalent to maximizing $a$ for this class of theories. Moreover, we show 
that the gauge theory result precisely matches the holographic free energy of the 
supergravity solution, and provide a similar matching of the 
 VEV of a BPS Wilson loop operator.
\end{abstract}

\end{titlepage}

\pagestyle{plain}
\setcounter{page}{1}
\newcounter{bean}
\baselineskip18pt
\tableofcontents



\section{Introduction and summary}\label{SecIntroduction}

Superconformal field theories (SCFTs) form an interesting class 
of supersymmetric QFTs. They arise naturally at fixed points of renormalization 
group flows, and the additional symmetry provides  important constraints on the theory. 
SCFTs also arise on one side of the AdS/CFT correspondence, 
giving dual descriptions of quantum gravity in AdS space.

Since the advent of AdS/CFT \cite{Maldacena:1997re} there 
has been enormous progress in the understanding and construction 
of SCFTs, often through their embedding in string theory.
There are many classes of such constructions. One 
approach is via the low-energy effective field theories
on branes probing singular spaces. The geometry of the singularity 
determines the amount of supersymmetry, gauge group, matter content and interactions 
of this low-energy theory. 

For example, one can engineer a
four-dimensional $\mathcal{N}=1$ 
quiver gauge theory as the worldvolume theory on a stack of 
$N$ D3-branes probing a Calabi-Yau three-fold singularity $X$.  
When $X$ admits a conical 
 metric 
of the form $g_X = \diff r^2 + r^2 g_Y$, such gauge theories 
are expected to flow to (in general strongly interacting) 
SCFTs in the IR, with large $N$ type IIB supergravity duals of the form AdS$_5\times Y$. This is by now a well-established story. 
An important development on the field theory side was $a$-maximization \cite{Intriligator:2003jj}.
Every four-dimensional $\mathcal{N}=1$ SCFT has a conserved 
$U(1)_R$ symmetry, which for example determines 
the scaling dimensions of chiral primary operators. However, 
due to potential mixing with non-R Abelian flavour symmetries, in
general symmetry principles alone do not determine the superconformal $U(1)_R$. 
In \cite{Intriligator:2003jj} it was shown that the latter is determined uniquely 
as the local maximum of the \emph{$a$-function} $a=a(R)$. 
Here $a(R)$  is a certain cubic combination of 't Hooft anomalies, which may thus be computed in the UV theory. Evaluated on the superconformal 
R-symmetry the $a$-function is simply the $a$ central charge, which 
for theories with AdS gravity duals of the above type is 
related to the volume of $Y$ \cite{Henningson:1998gx, Gubser:1998vd} :
\bea\label{aintro}
a &=& \frac{\pi^3}{4\, \mathrm{Vol}(Y)}N^2~.
\eea
This allowed for precision tests of the AdS/CFT correspondence, notably 
for the infinite family of $Y^{p,q}$ models in \cite{Gauntlett:2004yd, Benvenuti:2004dy}.
This work led to many related developments and generalizations.

There is a parallel, but more recent, story for three-dimensional SCFTs, starting 
with the seminal work of \cite{Aharony:2008ug}. Here one 
can for example engineer an $\mathcal{N}= 2$ superconformal theory 
on the worldvolume of $N$ M2-branes probing a Calabi-Yau four-fold 
singularity. These have large $N$ M-theory duals of the form AdS$_4\times Y$, 
with the UV gauge theory typically being a Chern-Simons quiver theory. 
Three-dimensional $\mathcal{N}=2$ superconformal field theories similarly 
have a conserved $U(1)_R$ symmetry, but there 
 is of course no central charge in three dimensions. However,
  it turns out that a closely analogous role is played by the free energy $F=-\log Z_{S^3}$, 
where $Z_{S^3}$ denotes the partition function on the three-sphere.  
For $\mathcal{N}=2$ theories this may be computed 
using localization techniques, and depends on a choice of R-symmetry for the UV theory 
 \cite{Jafferis:2010un}. The superconformal $U(1)_R$ locally maximizes the real part of $F=F(R)$
 \cite{Jafferis:2011zi, Closset:2012vg}.

In this note we show that there is a class of theories which link these two developments. 
We begin with a four-dimensional $\mathcal{N}=1$ ``parent'' quiver gauge theory, 
arising on $N$ D3-branes probing a Calabi-Yau three-fold singularity $X$. 
On T-dualizing/dimensional reduction one obtains a three-dimensional 
$\mathcal{N}=2$ theory on $N$ D2-branes probing $\R\times X$. We then add a Romans 
mass $F_0=n/(2\pi \ell_s)$, where $n$ is quantized to be an integer. 
This is well-known to generate a Chern-Simons coupling 
on the D2-branes, and we conjecture that the resulting theory flows 
to a superconformal fixed point in the IR, with very closely related properties to
the parent four-dimensional theory. In fact this construction 
was recently applied to $N$ D2-branes in flat space in \cite{Guarino:2015jca}, 
where the dual $\mathcal{N}=2$ AdS$_4\times S^6$ solution in massive IIA supergravity 
was constructed. For D2-branes probing $\R\times X$, where the 
conical singularity $X$ has link $Y=\{r=1\}$,
 this supergravity solution 
generalizes to AdS$_4\times M_6$, where $M_6\cong \mathcal{S}Y$ is topologically the 
suspension of $Y$. We find that the gravitational free energy of this solution is 
 \bea\label{freeintro}
 F_{\mathrm{gravity}}  &=& \frac{2^{1/3}\, 3^{1/6}\, \pi^3 }{5\, \mathrm{Vol}(Y)^{2/3}}\, n^{1/3}\, N^{5/3}~.
 \eea
The field theory free energy $F=F(R)$ localizes to a matrix model, 
and the large $N$ limit may be computed. Remarkably, we find 
 that to leading order at large $N$
\bea\label{Fa}
\Real F(R) &=& \frac{2^{5/3}\, 3^{1/6}\, \pi}{5}(nN)^{1/3} a(R)^{2/3}~,
\eea
where $a(R)$ is the $a$-function of the parent four-dimensional theory!
It immediately follows that locally maximizing $\Real F(R)$ is equivalent to locally 
maximizing $a(R)$. Moreover, substituting (\ref{aintro}) into (\ref{Fa}) 
shows that the field theory free energy agrees precisely with the 
supergravity result (\ref{freeintro}), for a generic Calabi-Yau 
three-fold singularity. We also compute the VEV of a certain 1/2 BPS Wilson loop operator $\mathcal{W}$
in the matrix model, finding the leading order large $N$ result
\bea
\Real \log \, \langle \, \mathcal{W}\, \rangle &=& \frac{2^{4/3}\, \pi}{3^{1/6}}\, (nN)^{-1/3}\, a^{1/3}~.
\eea
We show that this precisely agrees with minus the action of a dual fundamental string 
in AdS$_4$, where the string sits at either of the two points of suspension in the internal space
$M_6\cong \mathcal{S}Y$. 

The outline of this note is as follows. In section \ref{SecD2} we describe the string theory 
origin of the Chern-Simons gauge theories of interest, and write down the 
dual AdS$_4$ solution of massive type IIA supergravity. In section \ref{SecFT} 
we compute the large $N$ limit of the field theory matrix model, and 
compare to the supergravity results.


\section{D2-brane Chern-Simons theories}\label{SecD2}

\subsection{String theory set-up}\label{Secstring}

Consider a system of $N$ D2-branes on the background 
$\R^{1,2}\times \R\times X$, where $X$ is a local Calabi-Yau 
three-fold singularity. The metric on $\R\times X$ is given by
\bea\label{G2}
g_{\, \R\times X} &=& \diff z^2 + \diff r^2 + r^2 g_Y~,
\eea
where $z\in\R$, $r\geq 0$ and $(Y,g_Y)$ is a Sasaki-Einstein 
five-manifold. There are many constructions of 
such Calabi-Yau cones $X$, including infinite 
families of explicit metrics, as well as abstract 
existence results -- see \cite{Sparks:2010sn} for a 
review. Taking $Y=S^5$ with the round metric 
leads to flat space, while other simple examples 
include the homogeneous space $Y=T^{1,1}$ and the infinite family
$Y=Y^{p,q}$ \cite{Gauntlett:2004yd}. 

The low-energy effective theory on 
 $N$ D2-branes placed at $z=r=0$ is in general
 a three-dimensional  $\mathcal{N}=2$  field theory. 
When the singularity $X$ admits a 
Calabi-Yau (crepant) resolution, one 
expects this effective theory to be a quiver gauge theory, with superpotential.  
For example this is the case for toric Calabi-Yau singularities, 
for which the gauge group will be $U(N)^G$ where $G$ is the Euler number 
of the resolved space. Each copy of $U(N)$ arises as the gauge group on a 
fractional D-brane, wrapping various collapsed cycles at $r=0$, 
and the bifundamental matter fields in the quiver are 
massless strings between these branes. 
This set-up 
is perhaps more familiar in the context of $N$ D3-branes 
probing $X$, which leads to a four-dimensional $\mathcal{N}=1$ 
theory. A simple T-duality/dimensional reduction relates 
the two, and since the D3-brane theory will play a role 
later in the paper we shall refer to it as the ``parent'' theory. 

The Yang-Mills gauge coupling is dimensionful 
in three dimensions, but an alternative 
gauge kinetic term is provided by the Chern-Simons 
three-form. Suppose we have a three-dimensional $\mathcal{N}=2$ 
$U(N)^G$ quiver gauge theory, engineered as above.  
The $\mathcal{N}=2$ vector multiplet contains the gauge field $\mathcal{A}$, 
two real scalars $D$ and $\sigma$, and a two-component  spinor $\lambda$, 
all in the adjoint representation of the gauge group. 
Labelling the gauge groups by $a=1,\ldots,G$, we can consider 
adding the $\mathcal{N}=2$ Chern-Simons interaction
\bea\label{SUSYCS}
\mathcal{L}_{\mathrm{CS}} \ =\  \sum_{a=1}^G \frac{k_a}{4\pi} \int_{\R^{1,2}} \mathrm{Tr}\left[\mathcal{A}_a\wedge\diff\mathcal{A}_a+\tfrac{2}{3}\mathcal{A}_a\wedge\mathcal{A}_a\wedge\mathcal{A}_a+ (2D_a\sigma_a -\lambda_a^\dagger\lambda_a)\vol_3\right]~.
\eea
The Chern-Simons levels $k_a$
for a $U(N)$ or $SU(N)$ gauge group should be integer  in order for the theory to be gauge invariant.
Provided the sum of levels $k_a$ is zero, $\sum_{a=1}^G k_a=0$, such 
gauge theories generically describe the low-energy limit of 
$N$ M2-branes probing certain Calabi-Yau four-fold singularities. 
Such a relation to parent four-dimensional $\mathcal{N}=1$ 
gauge theories was first suggested in \cite{Martelli:2008si}, 
and later given a string theory derivation in  \cite{Aganagic:2009zk}.
The idea is that  M-theory 
on a Calabi-Yau four-fold singularity may often be reduced to type IIA string theory on a Calabi-Yau three-fold $X$
fibered over $\R$, with RR two-form flux arising as the curvature of the corresponding 
$S^1_{\mathrm{M}-\mathrm{theory}}$ bundle. The M2-branes reduce to $N$ D2-branes
probing $\R\times X$, whose $\mathcal{N}=2$ worldvolume theories 
are essentially just the dimensional reduction of the four-dimensional $\mathcal{N}=1$ 
D3-brane theory. Turning on the RR flux and fibering $X$ over $\R$
 induces Chern-Simons couplings in this theory, 
via the Wess-Zumino couplings on a D-brane.  

In this paper we would like to consider precisely the opposite type of Chern-Simons 
deformation, setting instead
\bea
k_a &=& k~, \qquad a=1,\ldots ,G~.
\eea
As explained in \cite{Gaiotto:2009mv}, in type IIA string theory this corresponds to adding
a Romans mass 
\bea\label{Romansmass}
F_0 & =& \frac{n}{2\pi \ell_s}~,
\eea
where $\ell_s$  denotes the string length and
\bea\label{ntok}
n &=& \sum_{a=1}^G k_a \ = \ G k~.
\eea
For $N$ D2-branes in flat space the low-energy effective 
theory is of course $\mathcal{N}=8$ $U(N)$ super-Yang-Mills, 
and turning on the Romans mass $n$ induces a Chern-Simons 
term via 
 the Wess-Zumino coupling
\bea
S_{\mathrm{WZ}} &=& (2\pi \ell_s^2)^2\mu_{\mathrm{D2}}\int_{\R^{1,2}} F_0 \cdot \frac{1}{2}
 \mathrm{Tr}\left(\mathcal{A}\wedge\diff\mathcal{A}+\tfrac{2}{3}\mathcal{A}\wedge\mathcal{A}\wedge\mathcal{A}\right) \nn\\ &= &  \frac{n}{4\pi}\int_{\R^{1,2}}\ \mathrm{Tr}\left(\mathcal{A}\wedge\diff\mathcal{A}+\tfrac{2}{3}\mathcal{A}\wedge\mathcal{A}\wedge\mathcal{A}\right)~,
\eea
where $\mu_{\mathrm{D2}}$ is the D2-brane charge. 
For our more general set-up of D2-branes at a Calabi-Yau singularity, a similar mechanism 
induces Chern-Simons couplings for each fractional brane, resulting in the relation (\ref{ntok}). 

Provided the Romans mass can be turned on  preserving $\mathcal{N}=2$ 
supersymmetry, the Chern-Simons terms above will be completed 
to the $\mathcal{N}=2$ couplings (\ref{SUSYCS}). 
At the level of the corresponding D2-brane 
supergravity solution this is not immediately clear, since 
turning on the Romans mass modifies the stress energy tensor 
and supersymmetry transformations, potentially sourcing 
other supergravity fields and deforming the D2-brane geometry. 
It would be interesting to try to construct such a solution explicitly, 
but we shall content ourselves in this note with instead writing down
the AdS$_4$ near horizon limit. We turn to this now.

\subsection{Dual supergravity solution}\label{Secgravity}

We would like to argue that the above D2-brane Chern-Simons 
theories flow in the IR to a superconformal fixed point with 
an AdS dual.  Such a relation between superconformal Chern-Simons gauge theories and
massive IIA was first suggested in \cite{Schwarz:2004yj}, and was recently reconsidered in 
\cite{Guarino:2015jca}. In the latter reference the authors considered 
$N$ D2-branes in flat space, together with an $\mathcal{N}=2$ Chern-Simons 
interaction generated by the Romans mass. The dual supergravity solution 
is a warped product AdS$_4\times S^6$. In our more general set-up,
topologically the near horizon geometry should 
be AdS$_4\times \mathcal{S}Y$, where $\mathcal{S}Y$ denotes the suspension of $Y$. 
This is because the link of the singularity $z=r=0$ in (\ref{G2}) 
is $z^2+r^2=1$, which has induced metric
\bea\label{sinecone}
\diff s^2_{\mathcal{S}Y} &=& \diff\alpha^2 + \sin^2\alpha\, \diff s^2_Y~,
\eea
where $r=\sin \alpha, z=\cos\alpha$. The metric (\ref{sinecone}) 
is called the \emph{sine cone} over $Y$, and is itself an Einstein metric 
admitting a Killing spinor. There are isolated 
 conical singularities at $\alpha=0$, $\alpha=\pi$, inherited 
 from the line of singularities at $r=0$ parametrized by $z$. 
 However, as mentioned in the last section, we expect the 
 $\mathcal{N}=2$ preserving Romans mass deformation 
 to also source other supergravity fields, and for the metric 
 (\ref{sinecone}) to correspondingly be deformed, but with 
 the same topology. 
 
 $\mathcal{N}=2$ supersymmetric AdS$_4\times M_6$ 
 solutions to massive IIA supergravity have been constructed in the literature, 
 notably in \cite{Gaiotto:2009yz, Petrini:2009ur, Lust:2009mb}.
 In particular in the latter reference $M_6$ is constructed from a 
 generic Sasaki-Einstein manifold $(Y,g_Y)$. However, these solutions are globally 
 well-defined only when $Y$ is a regular circle bundle over a K\"ahler-Einstein 
 manifold $M_4$, so that $M_6$ is the total space of an $S^2$ bundle over $M_4$. 
 This is not the topology we want, and the restriction on $Y$ being regular is too strong.
   We conjecture that the relevant $\mathcal{N}=2$ AdS$_4$ supergravity 
 solution in massive IIA supergravity is the following solution, 
 constructed recently in \cite{Guarino:2015jca}: 
 \bea\label{SUGRAsol}
 g &=&  \ex^{2A} \left(  g_{\mathrm{AdS}_4}
+ \frac{3}{2} \diff\alpha^2 
+   \frac{ 9 \sin^2 \alpha}{ 5 + \cos 2\alpha } \eta^2 +  \frac{ 6 \sin^2 \alpha}{ 3 + \cos 2\alpha }g_T  \right)~,\nn\\
\ex^{\Phi} &=&  \ex^{\Phi_0} \frac{  \left( 5 + \cos 2\alpha \right)^{3/4} }{ 3 + \cos 2\alpha }~,\qquad 
F_0 
\ = \  \frac{\ex^{-5\Phi_0/4}}{\sqrt{3}L}~,\nn\\
B &=& 6\sqrt{2}\, L^2\ex^{\Phi_0/2}\diff\left(\frac{\cos\alpha}{3+\cos2\alpha}\right)\wedge \eta~,\nn\\
F_2 &=& -\sqrt{6}\, L\, \ex^{-3\Phi_0/4}\left[\frac{4\sin^2\alpha\cos \alpha}{(3+\cos 2\alpha)(5+\cos 2\alpha)}\omega_T 
- 3\frac{(3-\cos 2\alpha)}{(5+\cos 2\alpha)^2}\dd\cos\alpha\wedge \eta\right]~,\nn\\
A_3 &=& 6L^3 \ex^{-\Phi_0/4} \Gamma + 3\sqrt{3}\, L^3\ex^{-\Phi_0/4}\frac{7+3\cos 2\alpha}{(3+\cos 2\alpha)^2}\sin^4\alpha \, \omega_T\wedge \eta~,
 \eea
 where $\diff\Gamma=\mathrm{vol}_{\mathrm{AdS}_4}$ and
  the warp factor is
 \bea
\ex^{2A} &=&  L^2  \left( 3 + \cos 2\alpha \right)^{1/2} \left( 5 + \cos 2\alpha \right)^{1/8}~.
 \eea
 Here $g$ is the ten-dimensional Einstein frame metric, $\Phi$ is the dilaton, $B$ is the $B$-field, while 
 $F_2$ is the RR two-form flux and $A_3$ is the RR three-form potential. All Sasaki-Einstein 
 metrics take the form
 \bea
g_Y &=& \eta^2 + g_T~,
 \eea
 where $\eta$ is a contact one-form and $g_T$ is a transversely K\"ahler-Einstein metric. 
 The corresponding transverse K\"ahler form has been denoted $\omega_T$, and 
 $\diff\eta = 2\omega_T$. The AdS$_4$ metric $g_{\mathrm{AdS}_4}$ has unit AdS radius.
 Finally $L$ and $\Phi_0$ are constants, 
 that we shall determine shortly.
 
 The ten-dimensional metric  takes a warped product form 
 $g=\ex^{2A}(g_{\mathrm{AdS}_4}+g_{M_6})$, where
 \bea\label{M6}
 g_{M_6} &=& \frac{3}{2} \diff\alpha^2 
+   \frac{ 9 \sin^2 \alpha}{ 5 + \cos 2\alpha } \eta^2 +  \frac{ 6 \sin^2 \alpha}{ 3 + \cos 2\alpha }g_T~.
 \eea
Here $\alpha\in[0,\pi]$, and this metric may be compared to the sine cone (\ref{sinecone}). 
The topology is precisely $M_6=\mathcal{S}Y$, with $\alpha=0$ and $\alpha=\pi$ being
isolated conical singularities. However, compared to (\ref{sinecone}) the 
Sasaki-Einstein metric at fixed $\alpha\in(0,\pi)$ has been squashed, with the relative sizes 
of $\eta^2$ and $g_T$ varying along the polar direction $\alpha$.\footnote{These
are called $\eta$-Sasaki-Einstein metrics in the mathematics literature.}

Recall that the Romans mass $F_0$ is quantized as in (\ref{Romansmass}), while 
$N$ D2-branes source $N$ units of six-form flux over $M_6$. These
Dirac flux quantization conditions 
 lead to the unlikely expressions
\bea
L &=& \frac{ \pi \ell_s \, n^{1/24}}{2^{7/48}\, 3^{7/24}}\left(\frac{N}{\mathrm{Vol}(Y)}\right)^{5/24}~, \qquad 
\ex^{\Phi_0} \ = \ \frac{2^{11/12}}{3^{1/6}\, n^{5/6}}\left(\frac{N}{\mathrm{Vol}(Y)}\right)^{-1/6}~,
\eea
where $\mathrm{Vol}(Y)$ denotes the volume of the Sasaki-Einstein metric on $Y$.
 The holographic free energy is given by the effective 
 four-dimensional Newton constant, $F=\pi/(2G_N)$, which 
 we calculate as
 \bea\label{freehol}
 F_{\mathrm{gravity}} \ = \ \frac{16\pi^3}{(2\pi \ell_s)^8}\int_{M_6} \ex^{8A}\, \vol_{M_6} &=& \frac{2^{1/3}\, 3^{1/6}\, \pi^3 }{5\, \mathrm{Vol}(Y)^{2/3}}\, n^{1/3}\, N^{5/3}~.
 \eea
 Setting $Y=S^5$ equipped with its round metric, topologically $M_6=\mathcal{S}S^5\cong S^6$ 
 and $\mathrm{Vol}(S^5)=\pi^3$, and (\ref{freehol}) reduces to the result in \cite{Guarino:2015jca}. 
 
 The metric (\ref{M6}) has Calabi-Yau conical singularities at $\alpha=0$, $\alpha=\pi$, 
 for generic $Y$. Nevertheless, these singularities do not lead to any divergences 
 in the holographic free energy, and we believe the supergravity solution (\ref{SUGRAsol})
 is the correct gravity dual. More precisely, although one expects some stringy degrees 
 of freedom to be supported at the Calabi-Yau singularities, in addition to the 
 massive IIA supergravity fields, the  supergravity solution  captures the leading order large $N$ behaviour. 
  We shall confirm this in the next section, by computing the free energy directly in field theory.
 

\section{Field theory}\label{SecFT}

Let us now turn to the dual field theory. The three-dimensional $\mathcal{N}=2$ superconformal field theories of interest in this paper 
have UV descriptions as Chern-Simons quiver gauge theories. There is some number $G$ of $U(N)$ gauge groups, 
with equal  Chern-Simons couplings $k$, together with various chiral multiplets in bifundamental representations 
$(\mathbf{N},\mathbf{\bar{N}})$ of $U(N)_a \times U(N)_b$, and adjoint representations of $U(N)_c$, specified by the quiver diagram.  
We assign the matter fields R-charges $\Delta^{ab}$, $\Delta^{c}$ respectively, consistent with the superpotential having R-charge 2,
which is necessary in order 
to define the theory on $S^3$ \cite{Jafferis:2010un}. 
 The partition function for such theories is given by~\cite{Kapustin:2009kz, Jafferis:2010un, Hama:2010av}
\bea
Z_{S^{3}}  \, = \,  \frac{1}{ \left( N! \right)^{G}} \int  \prod_{a=1}^{G} \left[ \prod_{i=1}^{N} \frac{\diff \lambda^{a}_i}{2 \pi}\right] \exp\left[\frac{\ii k}{4 \pi} \sum_{i=1}^{N} \left( \lambda^{a}_i \right)^{2}\right]\prod_{i\neq j}^{N}\sinh^{2}\left( \frac{\lambda^{a}_i-\lambda^{a}_j}{2} \right) \ex^{- F_{\mathrm{matter}}}\, ,
\eea
where $\lambda^a_i$, $i=1,\ldots,N$, are the eigenvalues of $2 \pi \sigma_a$ and
 the matter part is determined by the precise quiver data. Here a single bifundamental chiral multiplet transforming in a representation $(\mathbf{N},\mathbf{\bar{N}})$ of $U(N)_a\times U(N)_b$ and with R-charge $\Delta^{ab}$ contributes as
\bea
 F^{ab}_{\mathrm{matter}} &=&-\sum_{i,j=1}^N \ell \left[ 1-\Delta^{ab}+\frac{\ii}{2 \pi}(\lambda^{a}_i - \lambda^{b}_j)  \right]\label{freehmbifmatrix} \,,
\eea
and a chiral multiplet transforming in the adjoint representation of $U(N)_c$ and with R-charge $\Delta^c$ contributes as
\bea
 F^{\mathrm{adj},c}_{\mathrm{matter}} &=&- \sum_{i,j=1}^N \ell \left[ 1-\Delta^{c}+\frac{\ii}{2 \pi}(\lambda^{c}_i - \lambda^{c}_j)  \right] \, .\label{freehmadjmatrix}
\eea
Here we have used the common definition
\bea
\ell(z) &=& - z \log \left( 1-\ex^{2\pi \ii z} \right) + \frac{\ii}{2}\left[\pi z^{2} + \frac{1}{\pi} \mathrm{Li}_2\left( \ex^{2\pi \ii z} \right)\right] - \frac{\ii\pi}{12} \,.
\eea


\subsection{Matrix model large $N$ limit}

We next compute the large $N$ limit of a rather generic such matrix model, following the saddle point 
method of \cite{Herzog:2010hf}. Based on numerical simulations in a variety of examples, 
including both non-chiral and chiral models, 
we conjecture the following leading order ansatz for the large $N$ saddle point eigenvalue distribution:
\bea \label{ansatzeigenv}
 \lambda_i^{a} &=& N^{\nu}\left( x_i + \ii y_i \right) ~.
\eea
Here $x_i, y_i\in \R$ are $O(1)$ in the large $N$ expansion, and
we shall determine the exponent  $\nu>0$ analytically later. Notice that crucially 
all the $U(N)$ gauge groups have the same behaviour, {\it i.e.} the 
right hand side of (\ref{ansatzeigenv}) is independent of $a=1,\ldots,G$.

Following \cite{Herzog:2010hf} to compute the large $N$ limit of the free energy, we define a density
\be
 \rho(x) \ = \ \frac{1}{N} \sum_{i=1}^{N} \delta(x-x_i) \,,
\ee
with support on the finite interval $[-x_\star, x_\star]$. In the continuum limit, $\rho(x)$ becomes an integrable function satisfying
\bea
 \int_{-x_\star}^{x_\star} \diff x \, \rho(x) & =& 1 \,.
\eea
Furthermore a discrete sum over eigenvalues converges to a Riemann integral
\bea
 \frac{1}{N} \sum_{i=1}^{N} & \longrightarrow & \int_{-x_\star}^{x_\star} \diff x \, \rho(x) \,
\eea
 in the continuum limit $N \rightarrow \infty$.

\subsubsection*{Classical contribution}

Given the large $N$ behaviour of the eigenvalues~\eqref{ansatzeigenv}, the classical contribution to  the large $N$ free energy
$F=-\log Z_{S^3}$
 for a theory with $G$ $U(N)$ gauge groups is
\bea
 F_{\mathrm{classical}} &=& -\frac{\ii n}{4 \pi} N^{1+2 \nu} \int_{-x_\star}^{x_\star} \diff x \, \rho(x) \left[2 \ii x y(x) + (x^2-y(x)^2)\right] + o\left( N^{1+2\nu} \right) \, ,
\eea
where $n=G k$.

\subsubsection*{Vector multiplet contribution}

A single vector multiplet  appears as
\bea\label{freevmmatrix}
F_\mathrm{vector} & = & - \sum_{i<j}^N 2 \log 2 \sinh\left( \frac{\lambda_i-\lambda_j}{2} \right) \,
\eea
in the matrix model. In order to obtain the continuum limit we use the expansion
\bea \label{expsinh}
 \log 2 \sinh z &=& \sgn\, [\Real z] \, z - \sum_{m\geq 1} \frac{\ex^{- \sgn\, [ \Real z] \, 2 m z}}{m} \,.
\eea
The first term will cancel against the chiral multiplet contribution for the models of interest (see below). The second term can be evaluated by a repeated application of integration by parts
\bea \label{intbypartslogic}
&&2 N^2 \int_{-x_\star}^{x_\star} \diff x' \int_{-x_\star}^{x'} \diff x \, \rho(x') \rho(x) \sum_{m\geq 1} \frac{1 }{m} \ex^{- N^\nu m  \left [ x'-x+\ii (y(x')-y(x))\right]}\nn\\
&& = \  2 N^{2-\nu} \int_{-x_\star}^{x_\star} \diff x'  \rho (x')\rho(x) \sum_{m\geq 1}  \frac{1}{m^{2}} \ex^{- N^\nu m  \left [ x'-x+\ii (y(x')-y(x))\right]} \bigg|_{-x_\star}^{x'} 
- 2 N^{2}  \int_{-x_\star}^{x_\star} \diff x' \cdot\nn\\
&&\int_{-x_\star}^{x'} \diff x\, \rho(x') 
\sum_{m\geq 1}\frac{1}{m} \left[\frac{N^{-\nu}}{m}  \rho' (x) + \ii \rho(x) y'(x) \right] \ex^{- N^\nu m 
\left[ x'-x+\ii (y(x')-y(x))\right]}~.
\eea
The contribution in the second line coming from $x=-x_\star$ is exponentially suppressed and only the term with $x=x'$ contributes to leading order. We again perform integration by parts on the last line in~\eqref{intbypartslogic} and make sure to only extract the leading order terms for large $N$ (in particular the $\rho'(x)$ term is subleading). Repeated application of this procedure leads to a geometric series
\bea
 2 N^{2-\nu} \int_{-x_\star}^{x_\star} \diff x'  \rho^2(x') \sum_{m\geq 1}\frac{1}{m^2} \left[ 1-\ii y'(x') -  y'(x') ^{2} + \cdots \right]\,,
\eea
which, using $\sum_{\ell \geq 1} \frac{1}{\ell^2} = \frac{\pi^2}{6}$, gives the leading order behaviour of a single vector multiplet\footnote{Recall that we are neglecting the linear term in (3.11), as it will cancel for the models considered in this paper.}
\bea
 F_{\mathrm{vector}} &=&N^{2-\nu} \frac{\pi^{2} }{3} \int_{-x_\star}^{x_\star} \diff x\, \frac{ \rho^2(x)}{1+\ii  y'(x)} + o(N^{2-\nu})\,.
\eea


\subsubsection*{Chiral multiplet contribution}

A bifundamental chiral multiplet between $U(N)_a$ and $U(N)_b$, with R-charge $\Delta=\Delta^{ab}$, contributes 
to the matrix model via~\eqref{freehmbifmatrix}. 
We use a convenient expansion of the $\ell$-function (see for example~\cite{Martelli:2011qj})
\bea \label{expansionell}
 \ell(z) &=&  \sgn \, [ \Imag z] \, \frac{\ii \pi}{2} \left( z^2-\frac{1}{6}\right) + \sum_{m\geq 1}\left(\frac{z}{m}+ \ii \frac{\sgn\,  [\Imag z]}{2 \pi m^2}\right) \ex^{ \sgn\,  \left[\Imag z\right] \, 2 \pi \ii m z} \,.
\eea
Substituting in our ansatz~\eqref{ansatzeigenv} and taking the continuum limit, the first term in the expansions leads to a contribution of
\bea
&&- N^{\nu} \sum_{i<j}^{N} (1-\Delta) \, \sgn (x_i-x_j) \left[ x_i-x_j+\ii (y_i-y_j)\right] \nn\\
&&\longrightarrow  - N^{2+\nu}(1-\Delta)  \int_{-x_\star}^{x_\star} \diff x' \rho(x') \int_{-x_\star}^{x'} \diff x \, \rho(x) \left[ x-x'+\ii (y(x)-y(x'))\right] \,.
\eea
This precisely cancels the contribution coming from the vector multiplets. More precisely, it cancels against the first term in the expansion~\eqref{expsinh} provided that
\bea \label{sccond}
 G & = & \sum_{I\, \in \, \text{matter fields}} (1-\Delta^{I}) \,,
\eea
where the sum is taken over all matter fields (bifundamental and adjoint) in the quiver and $\Delta^{I}$ are their respective R-charges. 
Equation (\ref{sccond}) is simply $\mathrm{Tr}\, R =0$, where the trace is over all fermions in the theory 
(compare to (\ref{trR}) below). Precisely the same constraint arose in the Chern-Simons quiver 
theories in \cite{Martelli:2011qj}, for essentially the same reason. The parent 
 four-dimensional $\cN=1$ superconformal field theory certainly satisfies (\ref{sccond}), as 
 it is implied by the vanishing of the NSVZ beta functions for all the gauge groups 
 (see for example \cite{Benvenuti:2004dw}). As in  \cite{Martelli:2011qj},
 $\mathrm{Tr}\, R=0$ is an additional constraint on the three-dimensional theories 
 under consideration.

Focusing on the second part in the expansion~\eqref{expansionell}, and using exactly the same integration-by-parts argument as for the vector multiplet, a careful analysis shows that
\bea
&& \sum_{i,j=1}^{N} \sum_{m\geq 1}  \frac{\ii\,\sgn [x_i-x_j]}{2 \pi m^2} \ex^{ \sgn  [x_i-x_j] \, 2 \pi \ii m \left[1-\Delta+\frac{\ii}{2 \pi}N^\nu (x_i +\ii y_i -( x_j+\ii y_j)) \right]}\nn\\
&&  \longrightarrow
 -  N^{2-\nu} \sum_{m\geq 1}\frac{ \sin 2 \pi m (1-\Delta) }{\pi m^3}\int_{-x_\star}^{x_\star} \diff x  \, \frac{\rho^{2}(x)}{1+\ii y'(x)} \,,
\eea
for large $N$. For the remaining part of the expansion we compute the large $N$ contribution
\bea
&& \sum_{i,j=1}^{N} \sum_{m\geq 1}  \frac{1-\Delta+\frac{\ii}{2 \pi}N^\nu \left[x_i +\ii y_i -( x_j+\ii y_j)\right]}{m} \ex^{ \sgn  [x_i-x_j] \, 2 \pi \ii m \left[1-\Delta+\frac{\ii}{2 \pi}N^\nu (x_i +\ii y_i -( x_j+\ii y_j)) \right]}\nn\\
&& \longrightarrow  N^{2-\nu}   \sum_{m\geq 1} \left[\frac{ 2 (1-\Delta) \cos 2 \pi m (1-\Delta)}{m^2}  -\frac{ \sin 2 \pi m (1-\Delta)}{ \pi m^{3}} \right]\int_{-x_\star}^{x_\star} \diff x \,  \frac{\rho^{2}(x)}{1+\ii y'(x)} \,. \nn\\
\eea
It remains to evaluate the Fourier series. For $\Delta \in [0,1]$ these are given by
\bea
 \sum_{m\geq 1} \frac{\sin 2\pi m \Delta}{m^{3}}   &=& \frac{\pi^3}{3}\Delta(1-\Delta)(1-2\Delta) \,,\nn\\
 \sum_{m\geq 1} \frac{\cos 2 \pi m \Delta}{m^{2}} &=& \pi^2 \left(\Delta-\frac{1}{2}\right)^2-\frac{\pi^2}{12} \,.
\eea

Putting everything together we arrive at the final form of the large $N$ chiral multiplet contributions\footnote{Again recall that we are neglecting the linear term in \eqref{expansionell}, since we showed that it will cancel for the models considered in this paper.}
\bea
  F^{ab}_{\mathrm{matter}}&=&  N^{2-\nu}\frac{\pi^2}{3}(1- \Delta^{ab}) \left[1-2\left(1- \Delta^{ab} \right)^2\right] \int_{-x_\star}^{x_\star} \diff x \,  \frac{\rho^{2}(x)}{1+\ii y'(x)} \,,\nn\\
 F^{\mathrm{adj},c}_{\mathrm{matter}} &=&  N^{2-\nu}   \frac{\pi^2}{3}(1- \Delta^{c}) \left[1-2 (1- \Delta^{c})^2\right] \int_{-x_\star}^{x_\star} \diff x\,   \frac{\rho^{2}(x)}{1+\ii y'(x)} \,,
\eea
for bifundamenal and adjoint chiral multiplets of R-charges $\Delta^{ab}$, $\Delta^{c}$, respectively. The contribution of an adjoint chiral multiplet is derived in exactly the same fashion.

\subsubsection*{Large $N$ free energy}

We consider a generic three-dimensional $\mathcal{N}=2$ Chern-Simons quiver theory, with $G$ $U(N)$ gauge groups and some number of bifundamental and adjoint chiral multiplets, giving the R-charge spectrum $\{\Delta^{I}:I  \in \text{matter fields} \}$. Given the expressions obtained above, the saddle point large $N$ free energy for such a model is obtained by extremizing
\bea\label{generalintegralsetup}
 F &=&  \frac{ n}{4 \pi} N^{1+2 \nu} \int_{-x_\star}^{x_\star} \diff x \, \rho(x) \left[2  x y(x) -\ii  (x^2-y(x)^2)\right]\nn\\
 && + N^{2-\nu} \frac{\pi^{2} }{3} \left\{ G \, + \sum_{I \, \in \, \text{matter fields}} (1-\Delta^{I})\left[1-2 (1-\Delta^{I})^{2} \right] \right\} \int_{-x_\star}^{x_\star} \diff x   \frac{\rho^{2}(x)}{1+\ii y'(x)} \nn\\
  &&+ \, o\left( N^{2-\nu} \right)\,.
\eea
In order to find a non-trivial saddle point 
both contributions have to be of the same order, which determines $\nu=1/3$. This precisely agrees with numerical simulations of 
a variety of models.

Noting that $y(x)$ is real, computing the leading order behaviour is now a simple variational problem, with the following general solution
\bea
 y(x) &=& \frac{1}{\sqrt{3}}x \,, \label{spy}\\
 \rho(x) &=&\frac{3\cdot 2^{2/3} \pi ^2 n^{1/3}  \left\{ G+ \sum_I (1-\Delta^{I})\left[ 1-2 (1-\Delta^{I})^{2} \right] \right\}^{2/3}-4 n x^2}{4 \sqrt{3} \pi ^3 \left\{ G+ \sum_I (1-\Delta^{I})\left[ 1-2 (1-\Delta^{I})^{2} \right] \right\}} \,, \label{sprho} \\ 
 x_\star &=& \frac{\sqrt{3} \pi \left\{ G+ \sum_I (1-\Delta^{I})\left[ 1-2 (1-\Delta^{I})^{2} \right] \right\}^{1/3} }{2^{2/3} n^{1/3} }\,.\label{spxstar}
\eea
Substituting these expressions back into~\eqref{generalintegralsetup}, we arrive at the final form of the large $N$ free energy
\bea\label{freeft}
F &=& 
\frac{3 \sqrt{3} \pi}{20 \cdot 2^{1/3}}\left( 1-\frac{\ii}{\sqrt{3}} \right) \left\{ G\, + \sum_{I \, \in \, \text{matter fields}} (1-\Delta^{I})\left[ 1-2 (1-\Delta^{I})^{2} \right] \right\}^{2/3} n^{1/3} N^{5/3} \nn\\
&&+ \, o(N^{5/3}) \,.
\eea

\subsection{Comparison to supergravity}

Let us compare this result to our supergravity prediction~\eqref{freehol}. To do so, we start by making an important observation. The part  of~\eqref{freeft} which depends on the quiver data can in fact be rewritten in terms of the  $a$-function
of the  four-dimensional parent theory.
 In particular, using the relation~\eqref{sccond}, which  as mentioned above is implied by the four-dimensional theory having vanishing beta functions, one finds that
\bea\label{reltoa}
G N^{2}+ \sum_{I \, \in \, \text{matter fields}} (1-\Delta^{I})\left( 1-2 (1-\Delta^{I})^{2} \right) N^{2}&=& \frac{64}{9} a \,.
\eea
Here $a$ is the $a$-function of the four-dimensional $\cN=1$ superconformal parent theory. It can be expressed as~\cite{Anselmi:1997am, Anselmi:1997ys}
\bea \label{afunctiondef}
 a & = & \frac{3}{32}\left( 3 \tr\, R^{3}- 5 \tr\,  R \right)\,,
\eea
where $R$ is a choice of $U(1)_R$ symmetry. In terms of the quiver data we have
\bea\label{trR}
\tr\, R^{\gamma} &=& G \dim U(N) + \sum_{I \, \in \, \text{matter fields}} \dim \mathcal{R}^{I} (\Delta^{I}-1)^{\gamma} \,,
\eea
where $\dim \mathcal{R}^{I}$ is the dimension of the respective matter representation with R-charge $\Delta^{I}$. 
(Notice then that (\ref{sccond}) is simply $\mathrm{Tr}\, R =0$.)
Given this remarkable relation, we can express the free energy~\eqref{freeft} in terms of the $a$-function
\bea\label{freefta}
F(R) &=&\frac{2^{5/3}  3^{1/6} \pi}{5} \left( 1-\frac{\ii}{\sqrt{3}} \right)(nN)^{1/3} a(R)^{2/3} + o(N^{5/3}) \,,
\eea
and we see that $F$-maximization is in fact \emph{equivalent} to $a$-maximization for the three-dimensional SCFTs considered here. 
Evaluating the $a$-function on the superconformal $U(1)_R$ gives the $a$ central charge. 
This in turn is related by AdS$_5$/CFT$_4$ to the volume of the internal space $\mathrm{Vol}(Y)$~\eqref{aintro}. 
Putting all this together, we see that the real part of the  large $N$ free energy precisely agrees with the
supergravity result~\eqref{freehol}.

We conclude this subsection with a brief  comment on the imaginary part of the free energy (\ref{freefta}). 
This  is well-defined only 
modulo $2\pi$, and so is effectively $O(1)$ in the large $N$ expansion. Its value is also 
 altered by adding a Chern-Simons term for a background
Abelian gauge field \cite{Closset:2012vg}. One therefore doesn't expect 
to be able to match it to a leading order supergravity calculation.

\subsection{Wilson loop}

We may decorate our three-dimensional field theory with Wilson loops. A BPS Wilson loop in a representation $\mathcal{R}$ of a single $U(N)$ gauge group is given by
\bea
 \mathcal{W}_\mathcal{R} &=& \frac{1}{\dim \mathcal{R}} \tr_\mathcal{R}\left\{\mathcal{P} \exp \left( \oint_\gamma \diff s\, (\ii \mathcal{A}_\mu \dot{x}^{\mu}+\sigma | x |) \right)\right\} \,,
\eea
where $\mathcal{P}$ denotes the path-ordering operator, $\mathcal{A}_\mu$ is the gauge field with $\sigma$ the corresponding scalar in the $\cN=2$ vector multiplet, and $x^{\mu} (s)$ parametrizes the Wilson line $\gamma \subset S^{3}$. In order to preserve some supersymmetry, $\gamma$  lies on a Hopf circle in $S^{3}$. In particular this gives rise to $1/2$ BPS Wilson loops in the three-dimensional theory.
The VEV of such a $1/2$ BPS Wilson loop can be computed by insertion of $\mathcal{W}_\mathcal{R}$ into the path integral, and so expressed in terms of the matrix model via localization~\cite{Kapustin:2009kz}. This amounts to 
an additional factor of 
\bea
\tr_\mathcal{R} \, \ex^{2 \pi \sigma}\,,
\eea
where  $\gamma$ has length $2 \pi$.

We  now focus on Wilson loops in the fundamental representation of a quiver with $G$ $U(N)$ gauge groups and some generic matter. In the large $N$ limit the VEV is then given by
\bea
\langle \,  \mathcal{W}_\mathrm{fund}\,  \rangle & \longrightarrow &  G N \int_{-x_\star}^{x_\star} \diff x\, \rho(x) \, \ex^{N^{\nu}(x+\ii y(x))} \,.
\eea
The leading order saddle point configurations in~\eqref{spy},~\eqref{sprho} and~\eqref{spxstar} are not affected by the addition of a Wilson loop, since it is subleading. The large $N$ VEV of a $1/2$ BPS fundamental Wilson loop is therefore simply given by
\bea
\log \, \langle\,  \mathcal{W}_\mathrm{fund}\,  \rangle &=& \left( 1+\frac{\ii}{\sqrt{3}} \right) x_\star N^{1/3} + o(N^{1/3}) \,.
\eea
Substituting for the saddle point configuration of $x_\star$ as written in~\eqref{spxstar}, we end up with
\bea\label{largeNwilson}
\log \, \langle \, \mathcal{W}_\mathrm{fund} \, \rangle &=& \frac{ 2^{4/3} \pi  }{3^{1/6}}  \left(1+\frac{\ii}{\sqrt{3}}\right) (n N)^{-1/3 }  a(R)^{1/3} \,,
\eea
where we have used equation~\eqref{reltoa}. Again appealing to  the relation~\eqref{aintro}  leads to a precise supergravity prediction.

\subsubsection*{Dual fundamental string}

On general grounds the $1/2$ BPS fundamental Wilson loop should map to a fundamental string in AdS$_4$ \cite{Maldacena:1998im}. 
This wraps a disk $\Sigma_2 \subset \text{AdS}_4$, where $\Sigma_2 \cong \text{AdS}_2$ has boundary $\partial \Sigma_2 = \gamma \subset S^{3}$. 
It is   natural for this to sit at one of the two suspension points in the internal space, with $\alpha = 0$ or $\alpha = \pi$. 
Computing the 
fundamental string action simply reduces to computing the volume of $\Sigma_2$ in the string frame metric. Converting from Einstein to string frame introduces an additional factor of the ten-dimensional dilaton, $\ex^{\Phi/2}$. Putting everything together, we find
\bea
S_{\mathrm{string}} &=& \frac{6^{1/8}}{\pi \ell_s^{2}}  \, \ex^{\Phi/2}|_{\alpha=0,\pi} L^{2} \,  \mathrm{Vol} (\Sigma_2) \,.
\eea
The divergent area of $\Sigma_2 \cong\text{AdS}_2$ is here renormalized via a local counterterm, namely 
the length of the boundary, leading to the standard result
\bea
\mathrm{Vol} (\text{AdS}_2) &=&- 2 \pi \,.
\eea
We hence find that
\bea
S_{\mathrm{string}} &=& - \frac{2^{2/3} \pi ^2}{3^{1/6} n^{1/3}\, \mathrm{Vol}(Y)^{1/3}} N^{1/3} \,.
\eea
As expected, this precisely agrees with minus the real part of $ \log \, \langle \, \mathcal{W}_\mathrm{fund}\,  \rangle $ as given in~\eqref{largeNwilson}, if we use the relation~\eqref{aintro} to express the $a$ central charge  in terms of the internal space  volume $\mathrm{Vol}(Y)$.


\subsection*{Acknowledgments}

\noindent 
We thank Paul Richmond for discussions at an early stage of this work.
The work of M.~F. is supported by ERC STG grant 306260.  J.~F.~S. is supported by the Royal Society. 


\end{document}